\documentclass[a4paper]{svmult}
\usepackage{times}
\usepackage[dvips]{graphicx,epsfig}
\usepackage{amsmath,amsfonts,amssymb}
\usepackage{cite,url}

\begin{document}

\mainmatter

\newcommand{\lsim}{\mathrel{\mathop{\kern 0pt \rlap
  {\raise.2ex\hbox{$<$}}}
  \lower.9ex\hbox{\kern-.190em $\sim$}}}
\newcommand{\gsim}{\mathrel{\mathop{\kern 0pt \rlap
  {\raise.2ex\hbox{$>$}}}
  \lower.9ex\hbox{\kern-.190em $\sim$}}}

\def\@fnsymbol#1{if case#1\hbox{}\or*\or\dagger\or\ddagger\or\mathcar''278\or\mathc
har''27B\or|\or**\or\dagger\dagger\or\ddagger\ddagger\else\@ctrerr\fi\relax}
\long\def\symbolfootnote[#1]#2{\begingroup%
\def\thefootnote{\fnsymbol{footnote}}\footnote[#1]{#2}\endgroup}
\long\def\myfootnote#1{\begingroup%
\def\thefootnote{(\arabic{footnote}) }\footnote{#1}\endgroup}
\long\def\letterfootnote[#1]#2{\begingroup%
\def\thefootnote{\alph{footnote}}\footnote[#1]{#2}\endgroup}

\title*{Dynamical simulation of heavy-ion collisions 
in the energy range from a few tens MeV/A to a few hundreds MeV/A}

\author{\underline{{M.V.~Garzelli}}\symbolfootnote[7]{ 
Invited talk presented at the 27th International Workshop 
on Nuclear Theory, Rila Mountains, Bulgaria, June 23 - 28 2008.}
}

\titlerunning{Dynamical simulation of heavy-ion collisions} 
\authorrunning{{M.V.~Garzelli}}

\toctitle{Dynamical simulation of heavy-ion collisions in the 
energy range from a few tens MeV/A to a few hundreds MeV/A}

\tocauthor{{M.V.~Garzelli}}

\institute{{Dipartimento di Fisica, Universit\`a degli Studi di Milano \&
INFN, \\
via Celoria, 16 $\,\,$ I-20133 Milano, Italia\\
e-mail address: \texttt{garzelli@mi.infn.it}}
}

\maketitle

\begin{abstract}
The overlapping stage of 
heavy-ion reactions can be simulated by dynamical microscopical models,
such as those built on the basis of the Molecular Dynamics (MD) approaches,
allowing to study the fragment formation process. 
The present performances of the Quantum MD (QMD) code developed at the University of Milano are discussed, sho\-wing results concerning fragment and 
particle production at bombarding energies up to $\lsim$ 700 MeV/A,
as well as a preliminary analysis on the isoscaling behaviour of isotopic
yield ratios for reactions with isospin composition N/Z in the (1 - 1.2) range, 
at a 45 MeV/A bombarding energy.
\end{abstract}

\section{Introduction}

In the phase diagram of nuclear matter 
two different regions, where signals related to the probable 
onset of phase transitions have been detected,
have been explored with increasing interest. 
First of all, a phase-transition is expected
to occur at a critical temperature of a few MeV 
and a critical density relatively low compared to
the normal nuclear matter density. Subsaturation densities 
can be accessed by heavy-ion collisions in the intermediate energy domain 
(30 MeV/A - 200 MeV/A), which allow to study 
the thermodynamic properties of strongly interacting nuclear matter. The characteristic features of nuclear forces (long-range attraction and short-range repulsion) could be responsible of the so called liquid-gas phase transition in this region. Symmetry energy effects and Coulomb effects are recognized to play an important role, whereas the role of the nucleon spin has still to be investigated~\cite{phase}. Reactions in this energy domain are currently studied by means of a number of facilities worldwide. In particular, we mention the experiments performed by the INDRA Collaboration at Ganil, by the ALADIN Collaboration at GSI, and by the CHIMERA and ISOSPIN Col\-la\-bo\-ra\-tions at the LNS in Catania. 
On the other hand, re\-la\-ti\-vi\-stic and ul\-tra\-re\-la\-ti\-vi\-stic energy heavy-ion collisions can be used to explore a region at much higher ($\rho$, T) where another phase transition 
probably occurs: the one from hadronic matter made of baryons and mesons to the quark-gluon phase, already investigated at RHIC, and which will be further explored by the LHC heavy-ion program. 
According to lattice calculations,
the critical temperature for the last phase transition
or sudden cross-over into a deconfined quark-gluon plasma, 
is expected to be $\approx$ 180 MeV. 
In this paper we will focus on the first kind of reactions, with 
projectile energies in the range from tens MeV/A to a few hundreds MeV/A.

Among the possible signatures of a liquid-gas phase transition, we mention in
particular multifragmentation, i.e. the simultaneous breakup of an 
excited nuclear sy\-stem in a large number of IMF 
(Intermediate Mass Fragments, 3 $\le$ Z $\le \mathrm{Z_{max}}$, 
where $\mathrm{Z_{max}}$ is well below the total charge
of the nuclear system) (liquid phase), coexisting in case with LMF
(Light Mass Fragments, Z $<3$) (gas phase).   
Multifragmentation is a universal phenomenon observed in intermediate-energy 
nuclear interactions induced by 
hadrons/photons/heavy-ions~\cite{bondorf}. 
During the overlapping stage of heavy-ion collisions (typical time $\approx$ 
100 - 200 fm/c), depending of the projectile ion bombarding energy and impact parameter, the nuclear system can undergo compression and reach high excitation energies. As a consequence, it starts to expand and can go on expanding down to sub-saturation densities ($\rho \approx  0.1 - 0.3 \, \rho_0$, where $\rho_0$ is the normal nuclear density) and reach temperatures T $\approx 3 - 8$ MeV, where it becomes unstable and breaks up into multiple fragments. These conditions are ty\-pi\-cal of the liquid-gas mixed phase~\cite{botvina}.

One of the issues still open is if a thermal equilibration is reached in these 
reactions. The statistical description of multifragmentation relies on this assumption. Accor\-ding to this approach, the emissions 
from multifragmenting sources are not affected by the dynamics which have led 
to source formation. Only the source global thermodynamical properties are
important to determine the fragment pattern.  
The dynamical description of multifragmentation instead does not rely on this assumption. According to the dynamical approach, multifragmentation is a fast process, and the IMF detected as reaction products   
are related to nucleon-nucleon correlations 
already occurring during the ion-ion overlapping stage 
and surviving the following expansion phase (memory effects)~\cite{aiche07}. 

The fast stage of heavy-ion collisions can be simulated by means of Monte Carlo dynamical models, such as the 
Quantum Molecular Dynamics ones (QMDs), 
al\-lo\-wing 
to describe phase-space fluctuations in the ion-ion overlapping stage, which lead to fragment formation~\cite{aichevecchio}. 
These are microscopical models, at nucleon level.
Preferring a microscopical model to a macroscopical one is mainly motivated by
the fact that a large number of degrees of freedom can appear in a reaction. 
Depending on the bom\-bar\-ding energy of the projectile system and 
on the impact parameter, one has to consider that 
excitation and deformation of projectile and target ions,
neck formation and brea\-king,
nucleon transfer,
different way of fragmenting the composite system, 
and many other effects can simultaneously play a role~\cite{giant}.
It could be very difficult to take into account so many degrees 
of freedom in a ma\-cro\-sco\-pi\-cal model. 
In a mi\-cro\-sco\-pi\-cal dynamical model, on the other hand, 
they can be described in a more natural, automatic and unified way.

At the end of the fast stage ( $\approx 10^{-22}$ s) of ion-ion interactions simulated by QMDs, excited fragments can be present. In order to reproduce experimental data concerning particle and fragment emission, also the de-excitation process has to be accounted for. Fragment de-excitation takes place on a time scale 
($\approx$ $10^{-21}$ - $10^{-15}$~s) far larger than the ion-ion overlapping process and is thus more precisely and faster described by other models. Anyway, 
some attempts to extend dynamical simulations up to a few thousands fm/c 
(i.e. $\lsim 10^{-20}$ s) have also been carried out with success 
by means of modified advanced versions of QMD (see e.g.~\cite{papa}).  

In this paper, the QMD code developed in Milano~\cite{cosparmv,ndmv} is used to simulate the overlapping stage of ion-ion collisions, leading to excited fragments (prefragments). Primary fragments are
identified at the end of each QMD event (t $\approx$ 200 - 250 fm/c) by applying a phase-space correlation algorithm to the nucleon distribution. Two particles are supposed to belong to the same cluster if their relative di\-stan\-ce 
and momentum 
are within fixed amounts. Our QMD code has been interfaced to the fragment 
de-excitation model already implemented into the general purpose FLUKA transport and interaction code~\cite{fluka1,fluka2,fluka3}. The evaporation/fission/Fermi break-up mo\-du\-le~\cite{trieste} available in FLUKA allows to describe primary fragment de-excitation on a statistical basis. 
General features of both codes have been presented in previous papers.
In this paper, results of simulation concerning particle and fragment emission at e\-ner\-gies within a few hundreds MeV/A are shown and compared to experimental data available in literature. 
In particular, the code has been applied both to the simulation of 
heavy-ion collisions up to $\lsim$ 700 MeV/A, as described 
in Section~\ref{particle} and~\ref{fragment}, and to the study of 
isospin effects at energies $\approx$ 50 MeV/A, typical of multifragmentation
events, as explained in Section~\ref{iso}. Our conclusions, with some reference
to practical applications too, are given in Section~\ref{conclu}.

\section{Particle production}
\label{particle}

As far as thin targets are concerned, double-differential neutron production
cross-sections for inclusive reactions at energies around a few hundreds MeV/A
have been presented in Ref.~\cite{cosparmv}. 
Target thickness has been neglected in our simulations. 

\begin{figure}[h]\centering
\includegraphics[width=58mm]{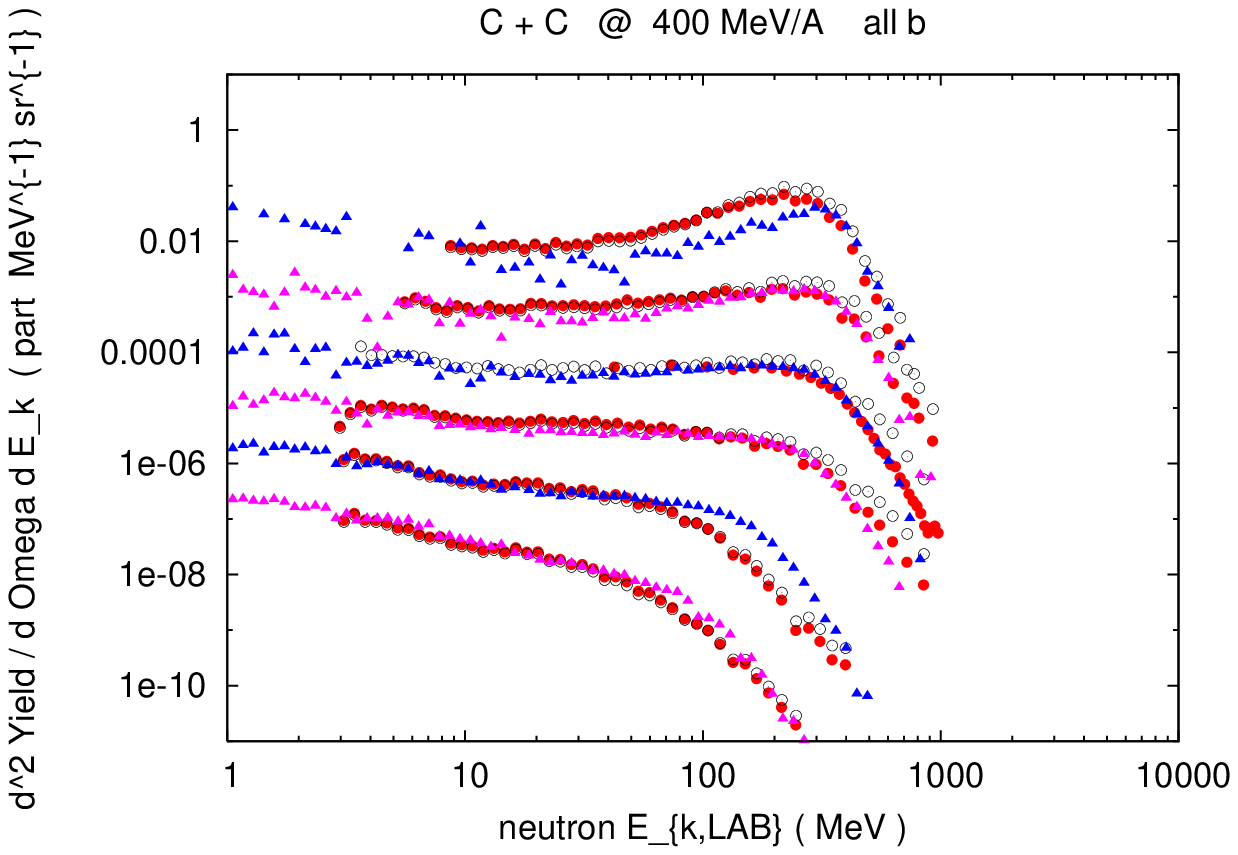}
\includegraphics[width=58mm]{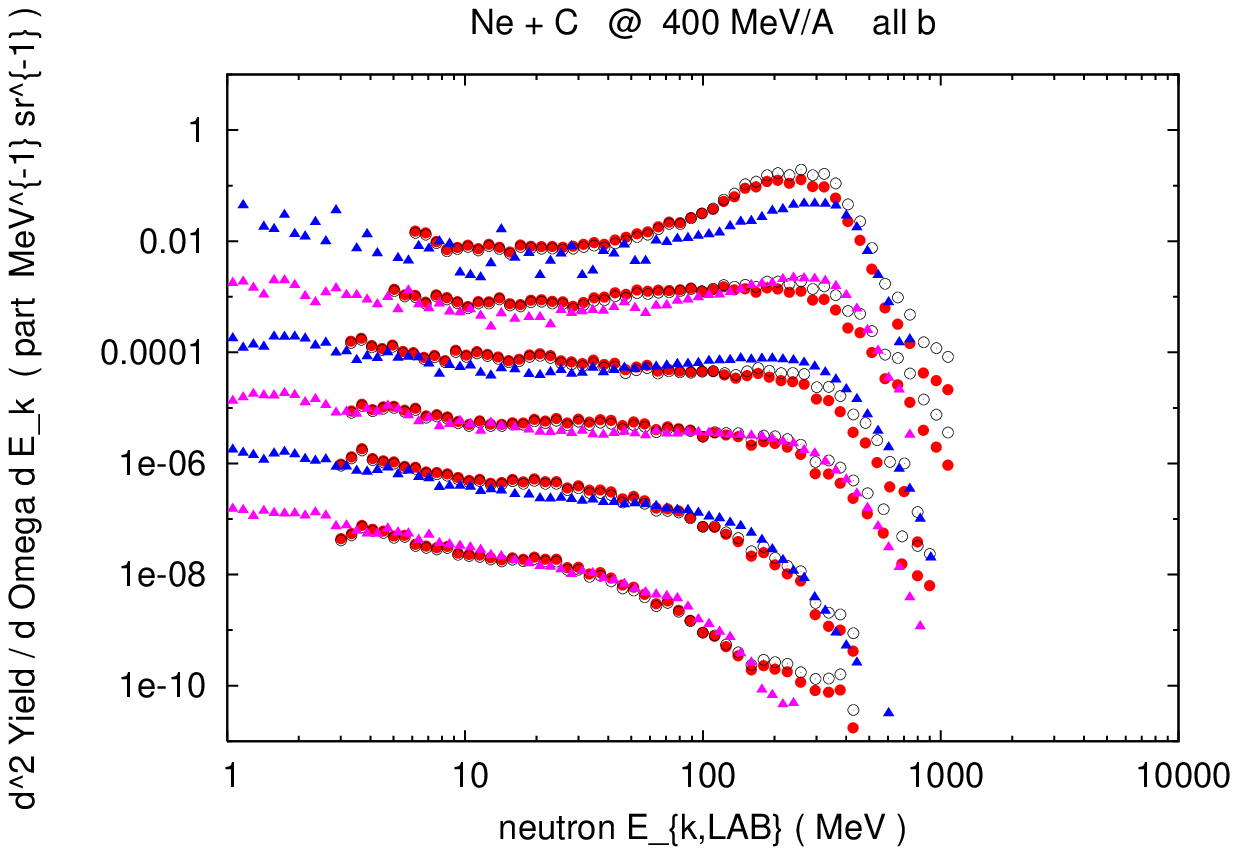}
\includegraphics[width=58mm]{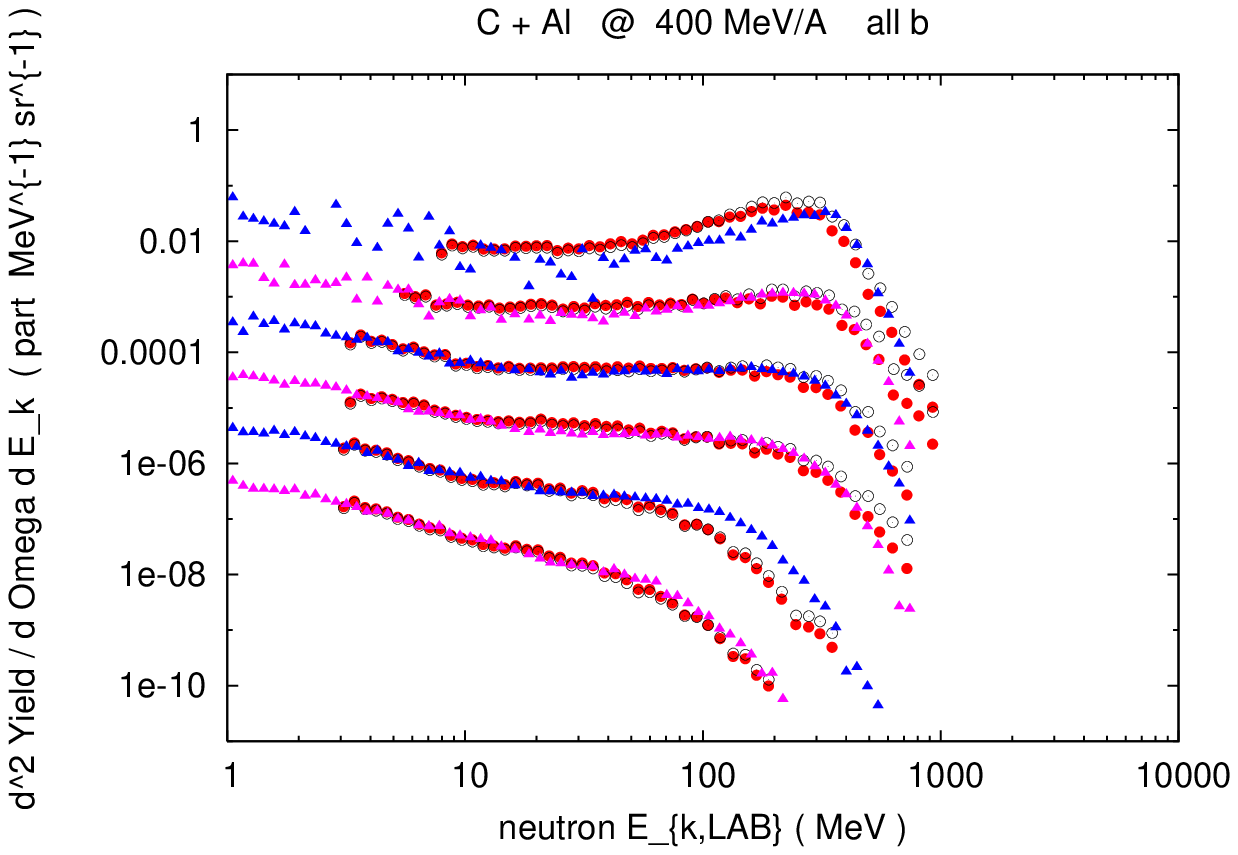}
\includegraphics[width=58mm]{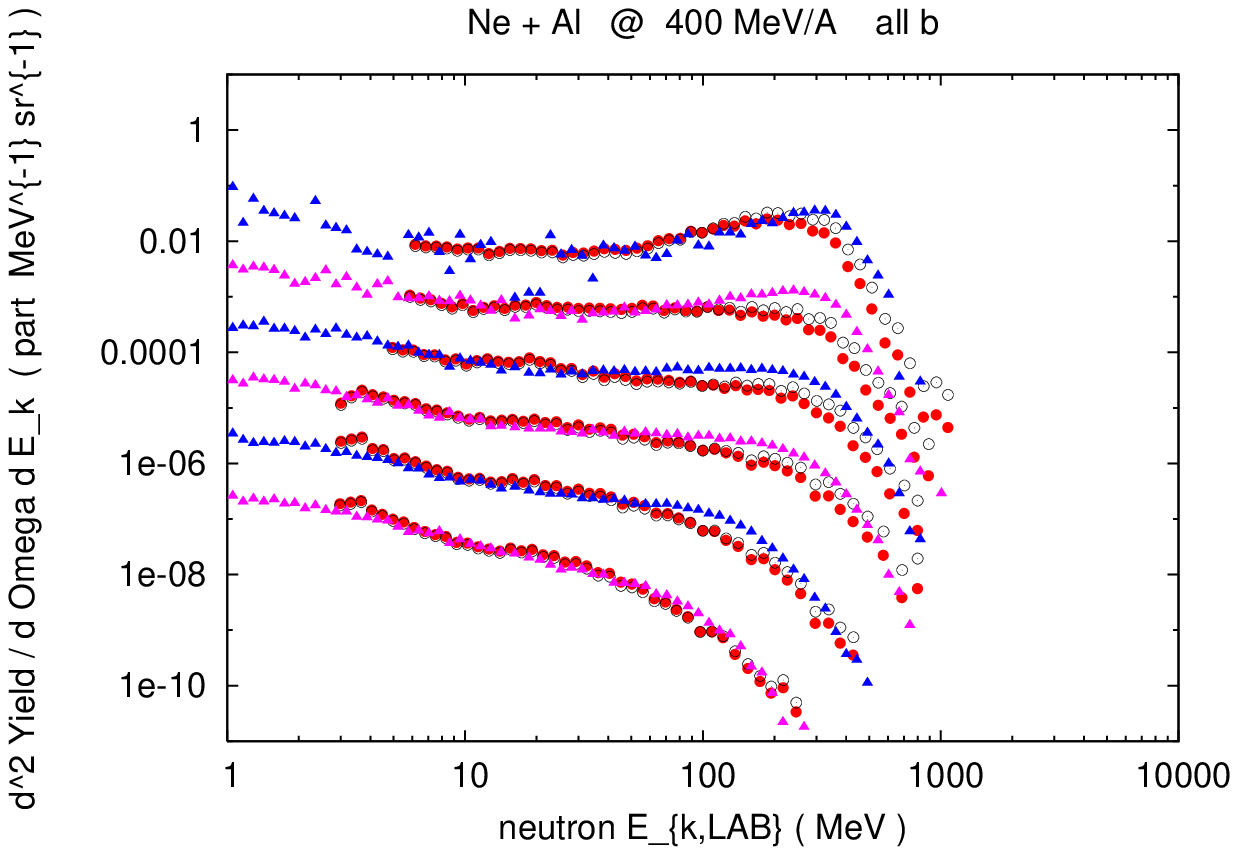}
\includegraphics[width=58mm]{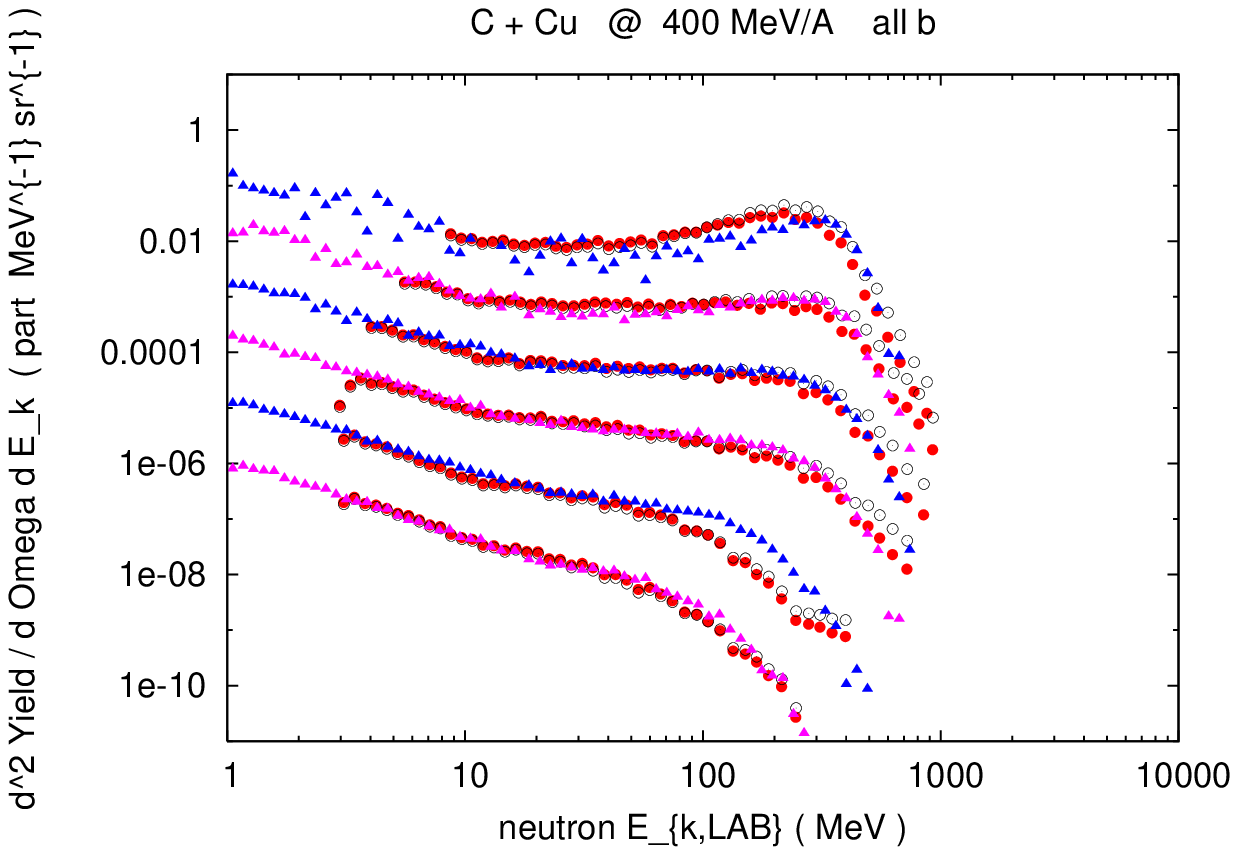}
\includegraphics[width=58mm]{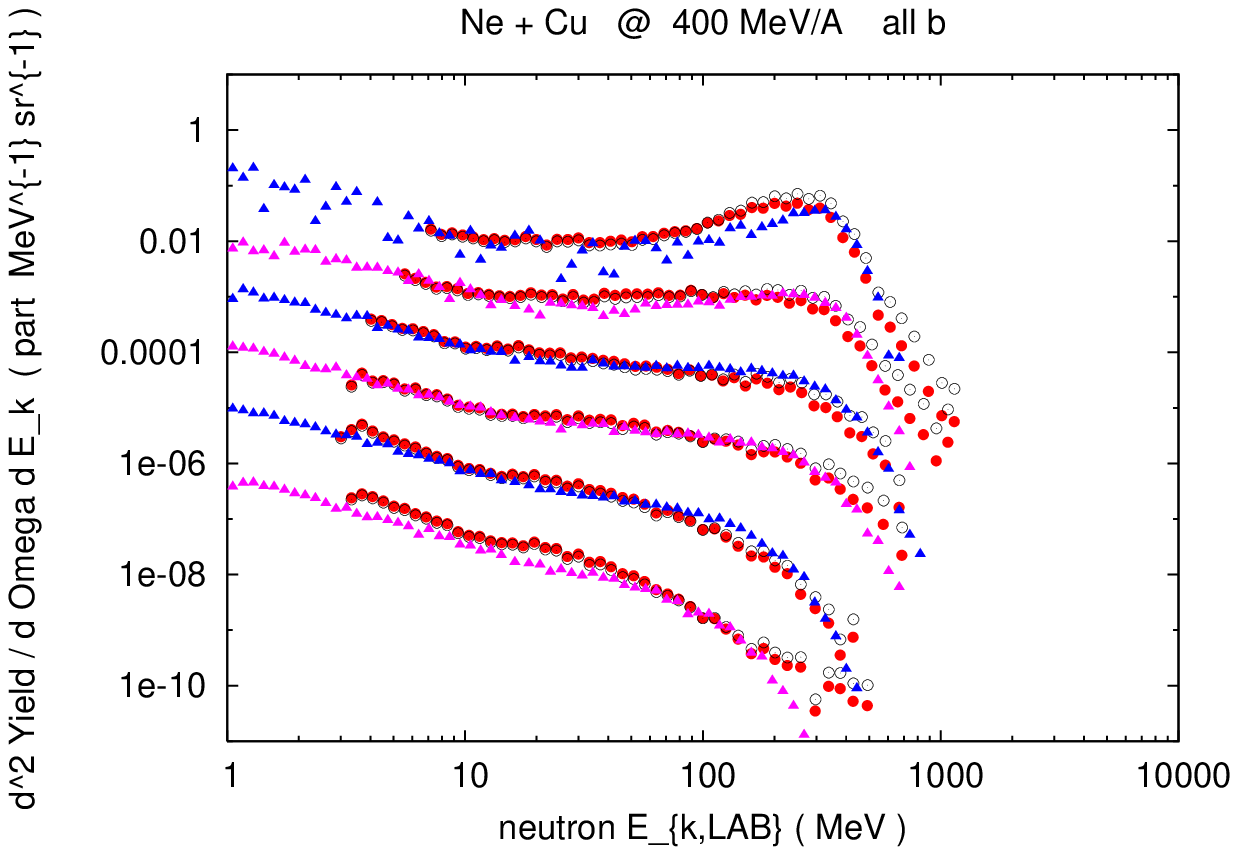}
\caption{\label{figsato} 
Double-differential neutron production yield
for C (left) and Ne (right) 
projectiles impinging on C (top), Al (center) and Cu (bottom) 
at a 400 MeV/A bombarding energy.
The results of the theoretical simulations made by QMD + FLUKA are shown by
solid triangles, whereas the experimental data re-evaluated 
by Satoh et al.~\cite{sato} are shown by filled circles.
The old experimental data~\cite{sato2} are shown by empty circles.  
In each panel, distributions
at $0^{\mathrm{o}}$, $7.5^{\mathrm{o}}$, $15^{\mathrm{o}}$, $30^{\mathrm{o}}$, 
$60^{\mathrm{o}}$, $90^{\mathrm{o}}$ 
angles with respect to the incoming
beam direction in the laboratory frame, have been multiplied by decreasing 
powers of 10, for display purposes.}
\end{figure}

As far as thick targets are concerned, a recent paper~\cite{sato} 
claimed that old data on neutron double-differential yields,
published in previous papers~\cite{kurosawacarbon,kurosawaneon} 
by the same working group, suffer
from an underestimation of the neutron detector efficiencies, with 
errors increasing with increasing neutron energy. 
The authors of~\cite{sato} have
thus re-evaluated the old results taking into account updated estimates
of the detection efficiencies and published new data from recent
measurements. 
We simulate the same reactions, taking into account target thicknesses, 
chosen in the experiment in such a way that the
energy losses of projectile ions lead to their stop inside the targets. 
The results of our theoretical simulations of neutron emission from 
C and Ne beams impinging on C, Al and Cu systems 
at a 400 MeV/A bombarding energy are shown
in Fig.~\ref{figsato} together with both 
the re-evaluated and the old experimental
data, where available. It is apparent that, even if, in general, 
the results of the theoretical simulations better agree 
with the re-evaluated data than
with the old ones as far as the tails at higher energies are concerned, there
are energies/angles where the theoretical simulations better agree
with the old data than with the re-evaluated ones.
In particular, in case of C projectiles, the results of the simulations
always overe\-stimate the experimental data
for the emission of high-energy neutrons at a $\sim 60^{\mathrm{o}}$ angle. 
The same trend at $60^{\mathrm{o}}$ has been retrieved by
using a completely independent code, e.g. the PHITS code~\cite{phits} 
including the JQMD model~\cite{jqmd}, as shown in ~\cite{sato}. 
These facts point out that, besides the uncertainties indeed present 
in the models for theoretical simulations, further uncertainties in the 
experimental data could exist.

\section{Fragment production}
\label{fragment}

\begin{figure}[h]\centering
\includegraphics[width=50mm]{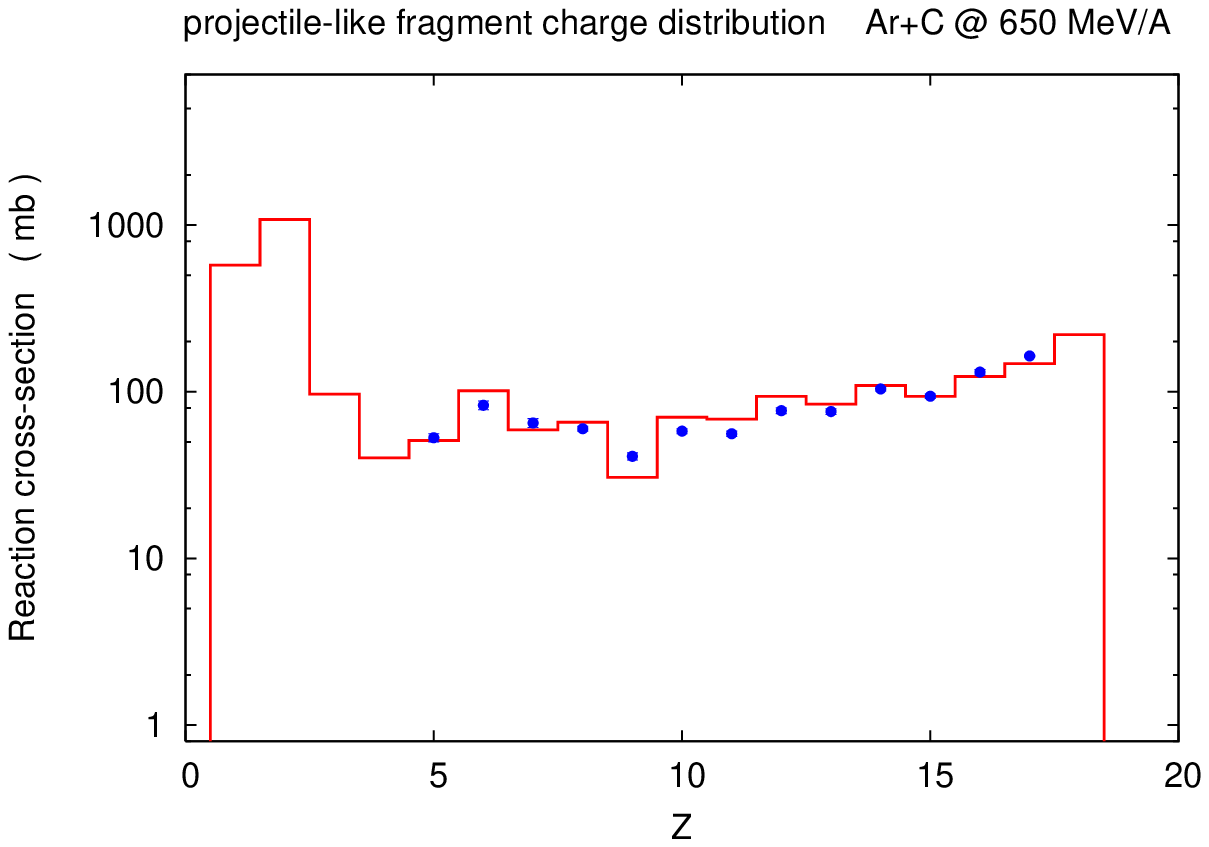}
\includegraphics[width=50mm]{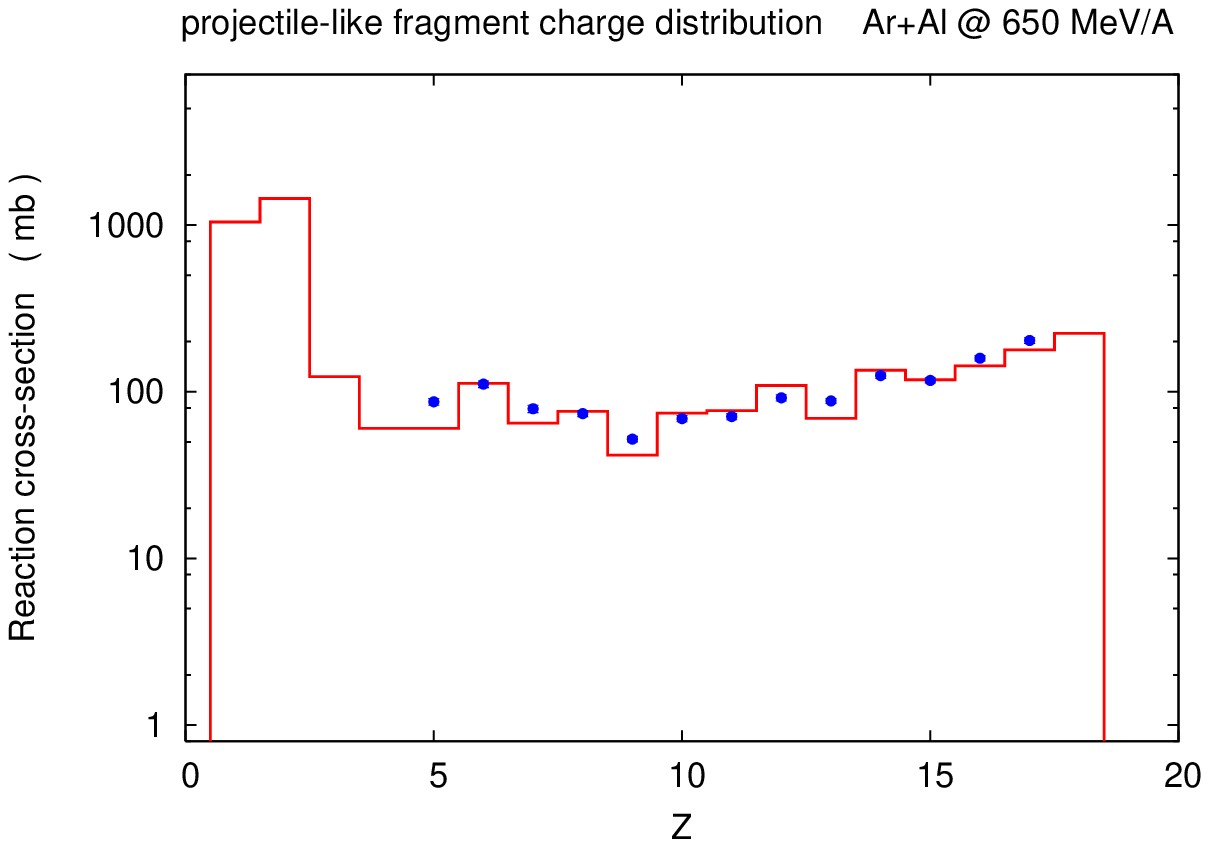}
\includegraphics[width=50mm]{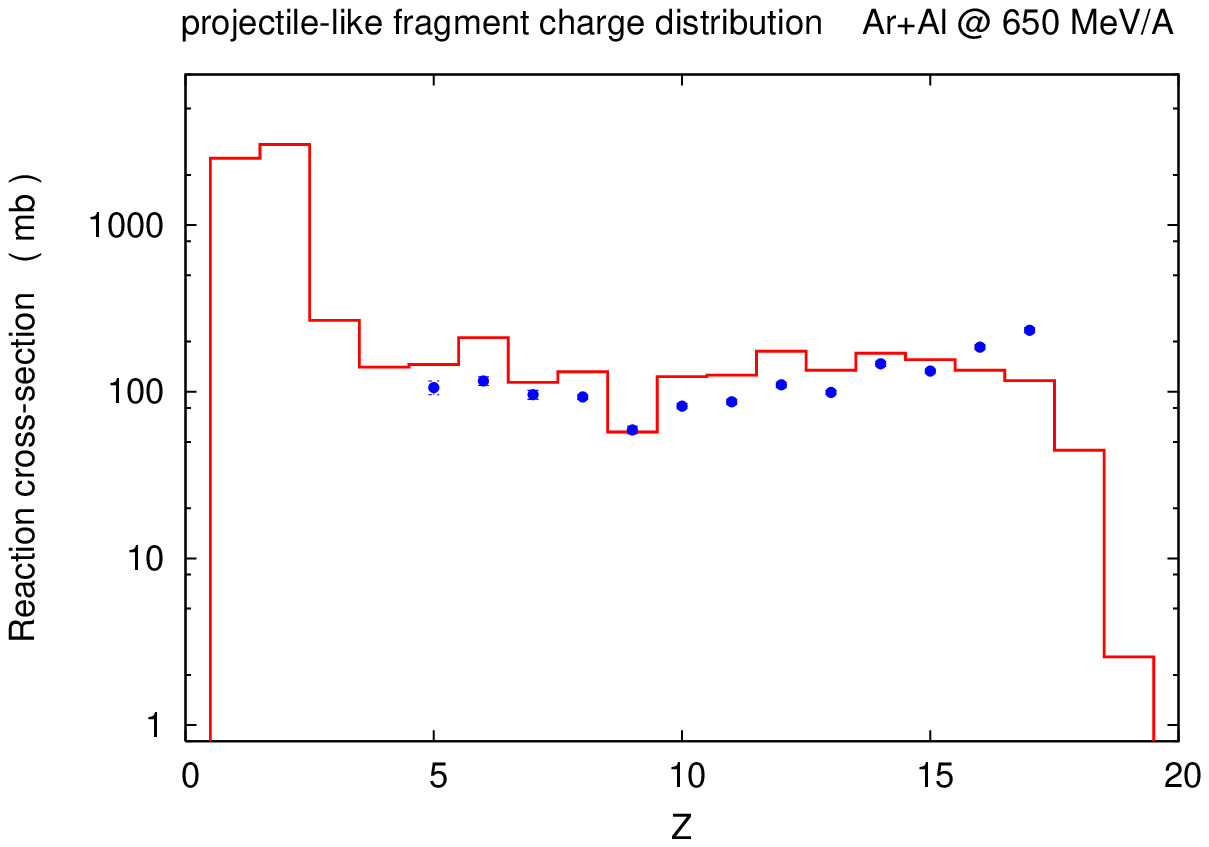}
\caption{\label{figar650} Projectile-like 
fragment production cross sections for 
650 MeV/A Ar beams fragmenting on C (top left), Al (top right) 
and Ar (bottom) targets,
calculated by QMD + FLUKA (solid histograms) in comparison with experimental
data taken from Ref.~\cite{zeitlinnuovo} (points). For Z~=~1 only the 
contributions to the cross-sections due to fragments (A $>$ 1) are shown.}
\end{figure}

\begin{figure}[h]\centering
\includegraphics[width=50mm]{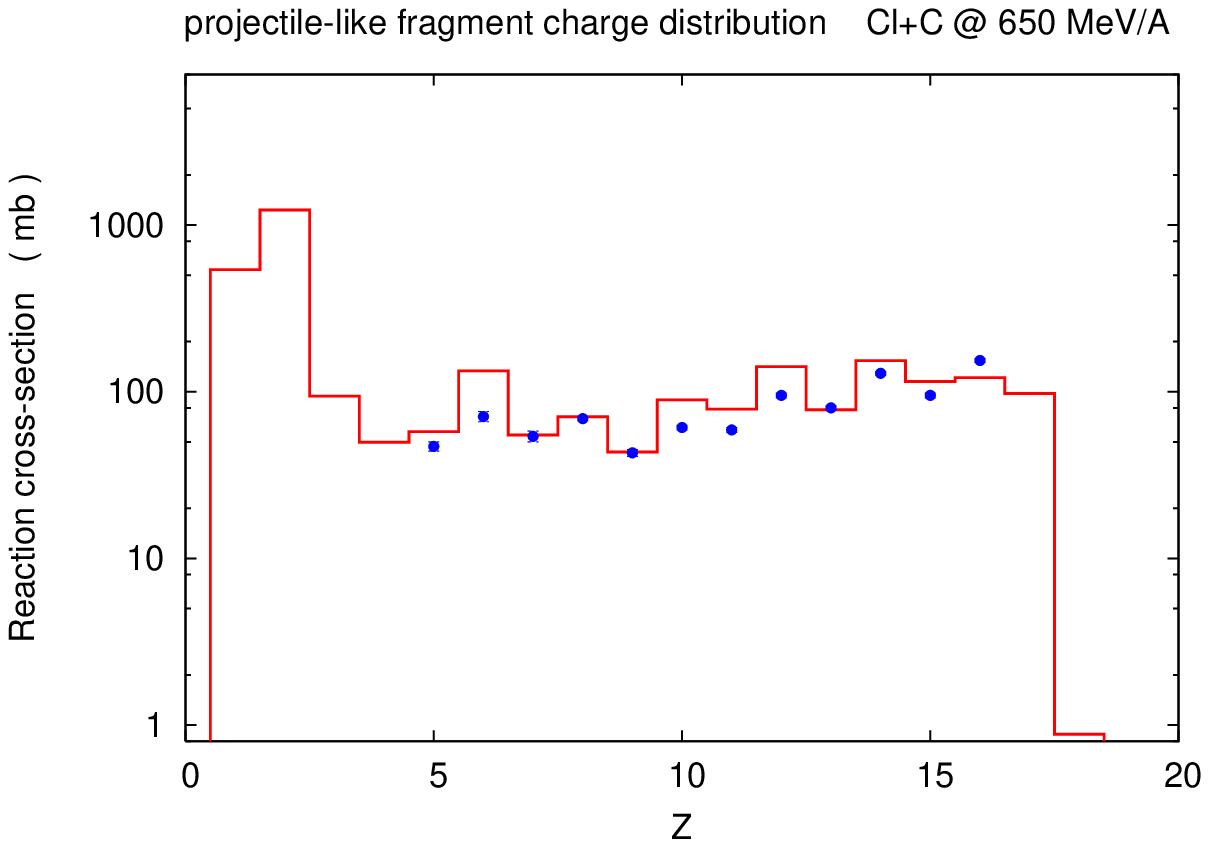}
\includegraphics[width=50mm]{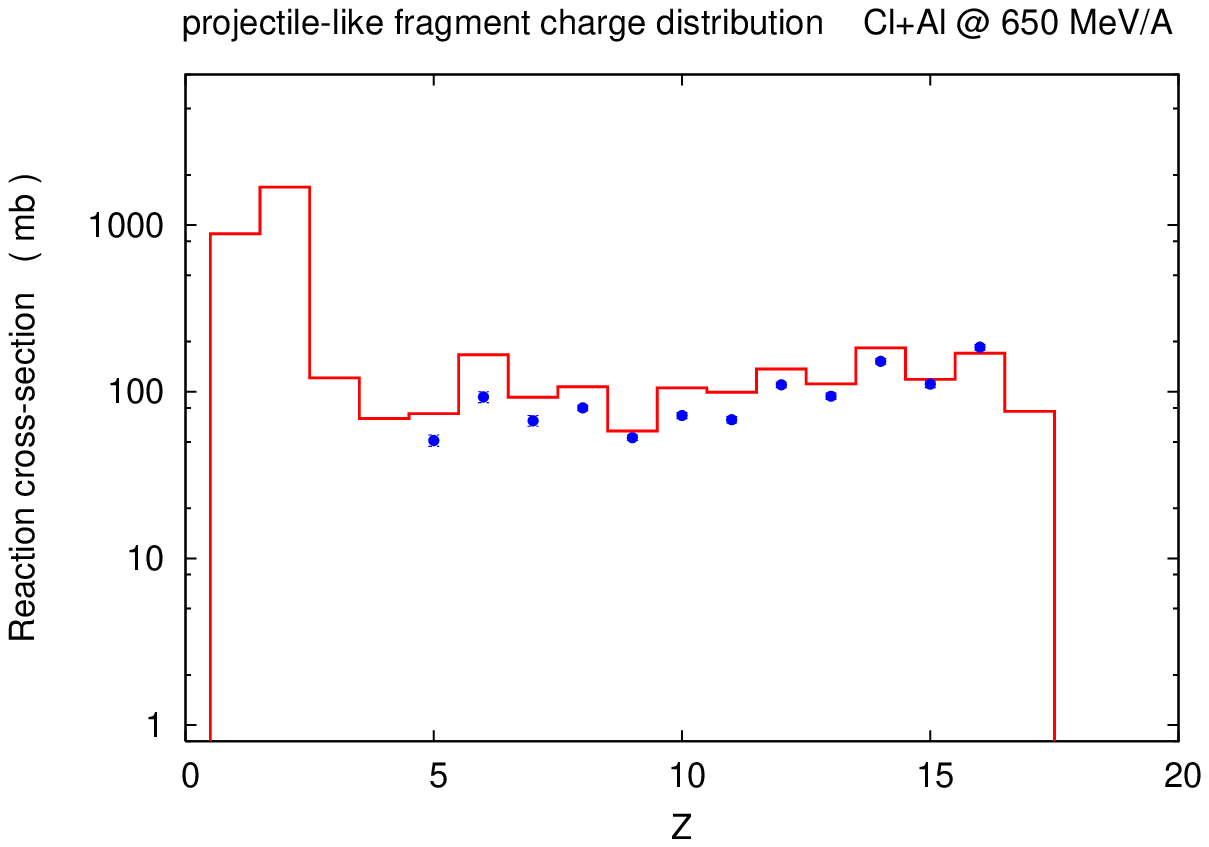}
\includegraphics[width=50mm]{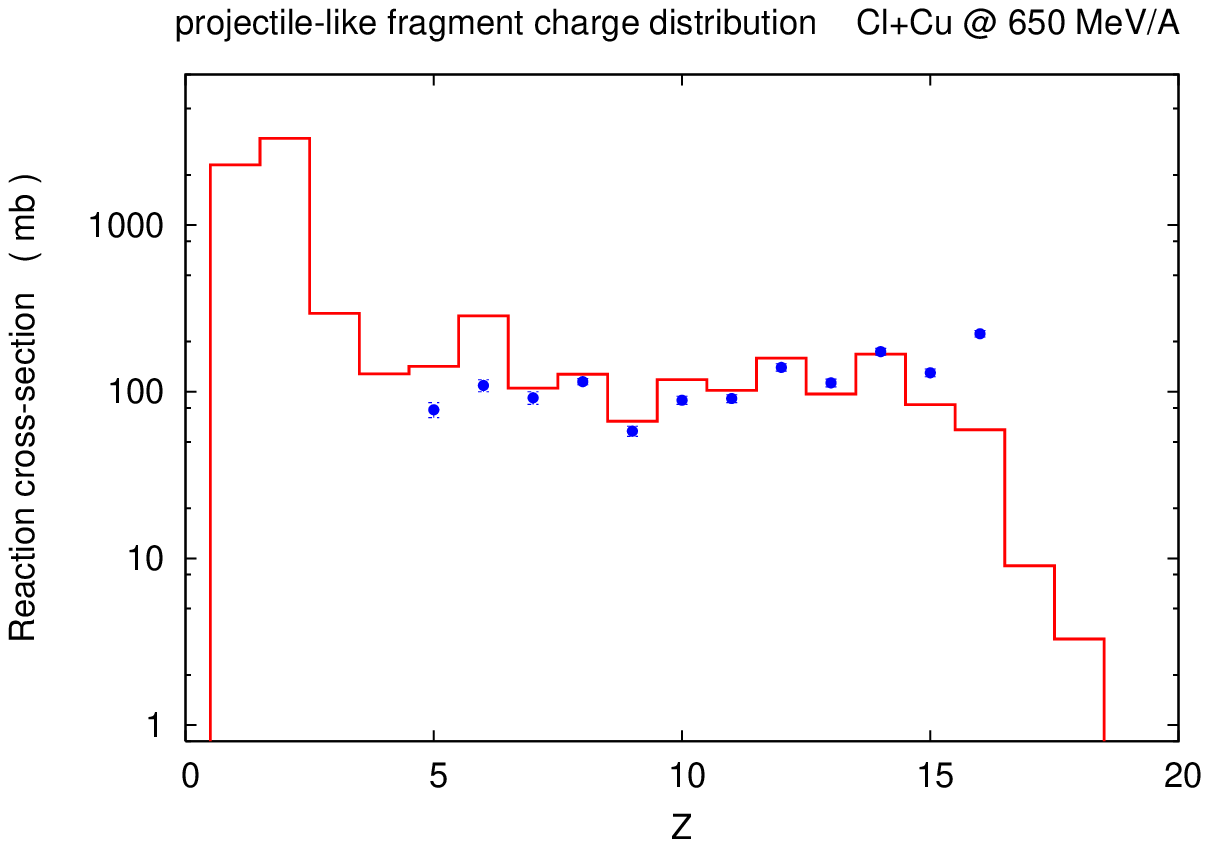}
\includegraphics[width=50mm]{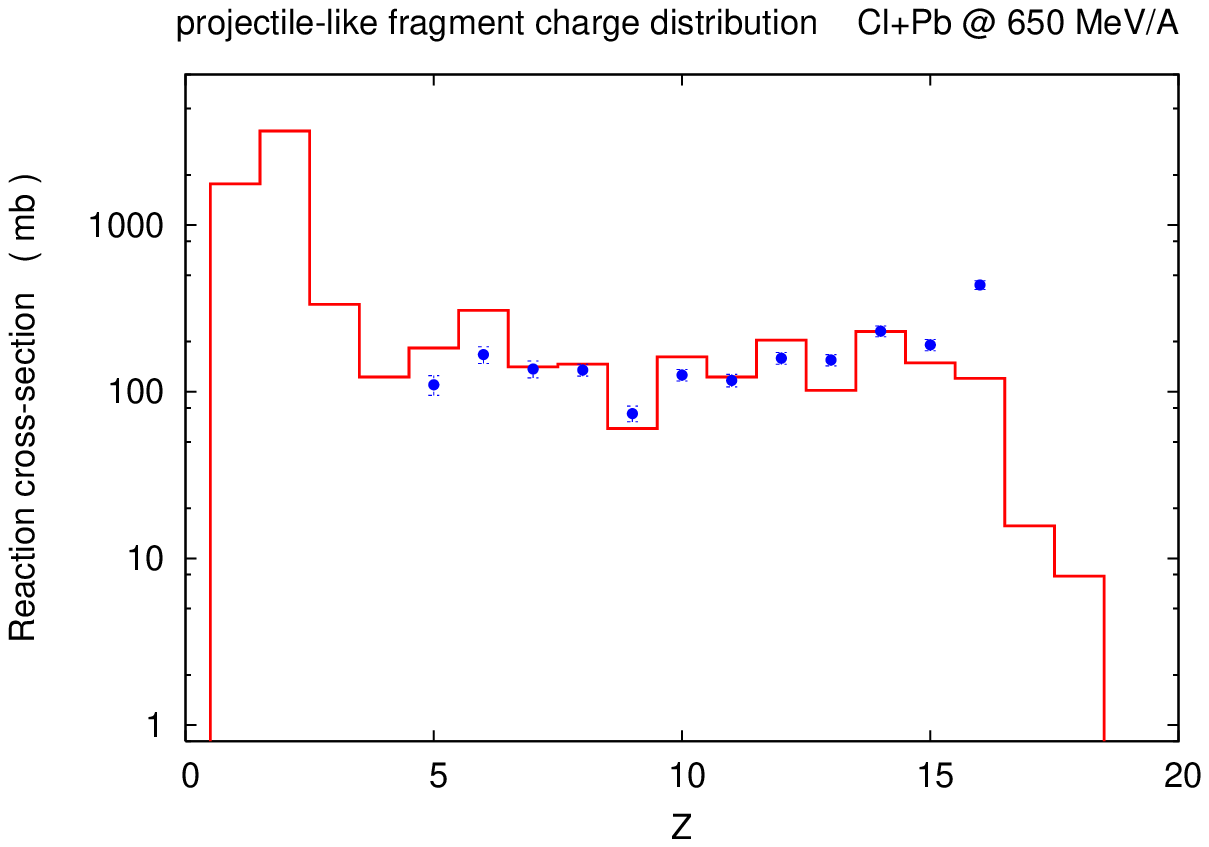}
\caption{\label{figclo650} Projectile-like 
fragment production cross sections for 
650 MeV/A Cl beams fragmenting on C (top left), Al (top right), 
Ar (bottom left) and Pb (bottom right) targets
calculated by QMD + FLUKA (solid histograms) in comparison with experimental
data taken from Ref.~\cite{zeitlinnuovo} (points). For Z = 1 only the 
contributions to the cross-sections due to fragments (A $>$ 1) are shown.}
\end{figure}

As far as fragment production is concerned, we calculate projectile-like 
fragment production cross-sections for Cl and Ar projectiles impinging on
C, Al, Cu and Pb targets at a 650 MeV/A bombarding energy.
Experimental data con\-cer\-ning these reactions were taken at 
the NSRL at the Brookeven National Laboratory, US and at
the HIMAC 
at the National Institute of Radiological Science, Japan and 
pu\-bli\-shed in Ref.~\cite{zeitlinnuovo}. 
The results of our simulations are shown together with the 
experimental data for Ar and Cl projectiles
in Fig.~\ref{figar650} and ~\ref{figclo650}, respectively.
Despite the overall reasonable agreement,
it is apparent that the theoretical simulations underestimate the production
of the most massive fragments, in case of the most massive targets. 
These fragments are emitted in very pe\-ri\-phe\-ral collisions.
A reason of this underestimation can be ascribed to electromagnetic 
dissociation, which increases with the system mass and
was not accounted for by our simulations. 
Anyway, we estimated that this effect 
is not enough to explain the discrepancies.
On the other hand, for lighter fragments, corresponding to less peripheral
collisions, the agreement between theoretical expectations and experimental
data is more reasonable and the odd-even effect observed in experimental data 
is reproduced by the simulation. In our simulations the last feature 
can be ascribed to de-excitation emission effects. For the lightest 
fragments which can be identified by the experiment the
agreement is sometimes slightly worser. The experimental procedure allows
charge identification by means of the $\Delta E$ deposited in silicon
detectors. With decreasing charge, the peaks in the (Z, count of events) plots
which allow charge identification become 
increasingly broader~\cite{zeitlinnuovo}. 
For $\mathrm{Z_{frag}} < \mathrm{Z_{primary}}/2$ it was not 
possible to distinguish the peaks corresponding to different charges with 
the large-acceptance ($\sim 5^{\mathrm{o}} - 13^{\mathrm{o}}$)
detectors used in the experiment. Thus, experimental data were obtained for 
fragments in the range $ 5  \le \mathrm{Z_{frag}} < \mathrm{Z_{primary}}/2$ 
by using small 
acceptance ($\sim 1^{\mathrm{o}} - 2^{\mathrm{o}}$) detectors. 
Furthermore, 
while the detectors measure 
fragmentation within a small angle around the beam axis,
the angular distributions of lighter fragments
are less peaked in forward direction. Thus, in our
opinion, the uncertainties 
in charge identification together with those
on the estimation of the  
correction due to the broadness of the angular distributions 
could largely affect the results on the cross-sections for light fragment 
production. 
 
\begin{figure}[h]\centering
\includegraphics[width=57mm]{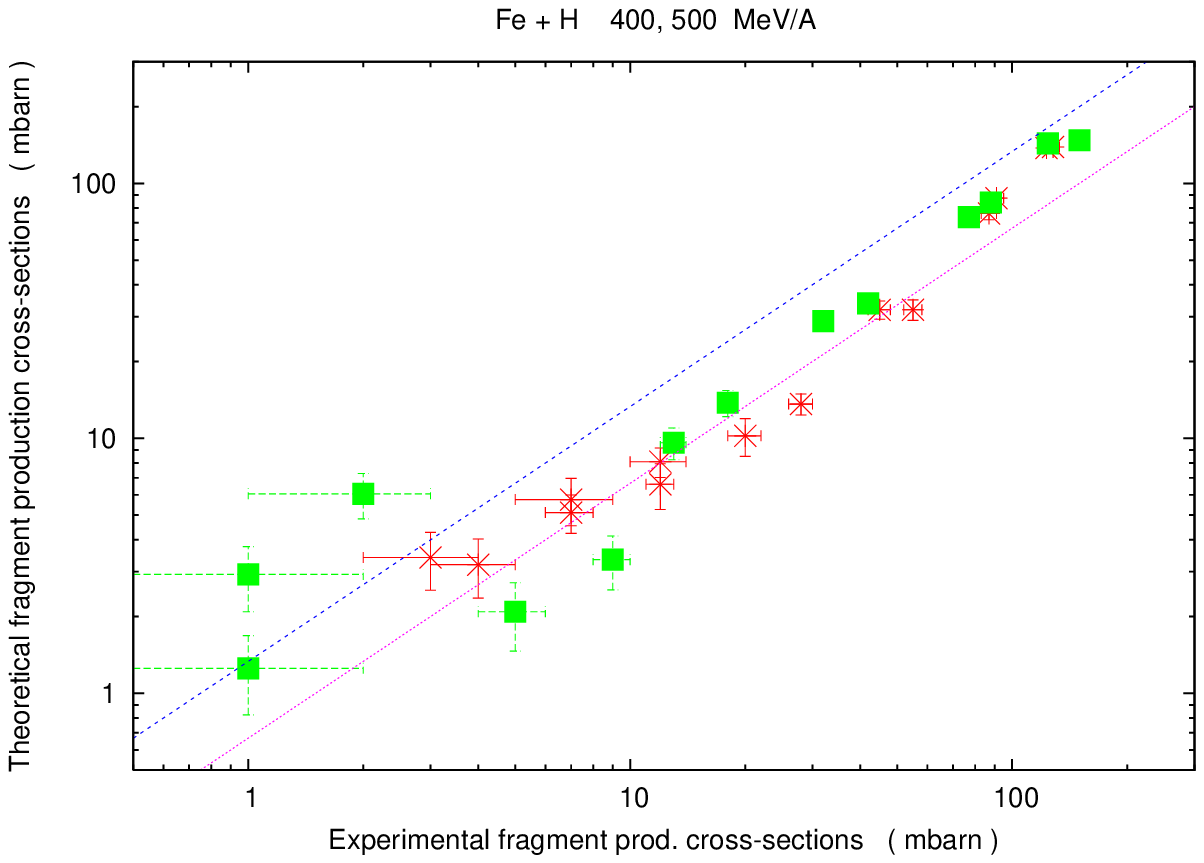}
\includegraphics[width=57mm]{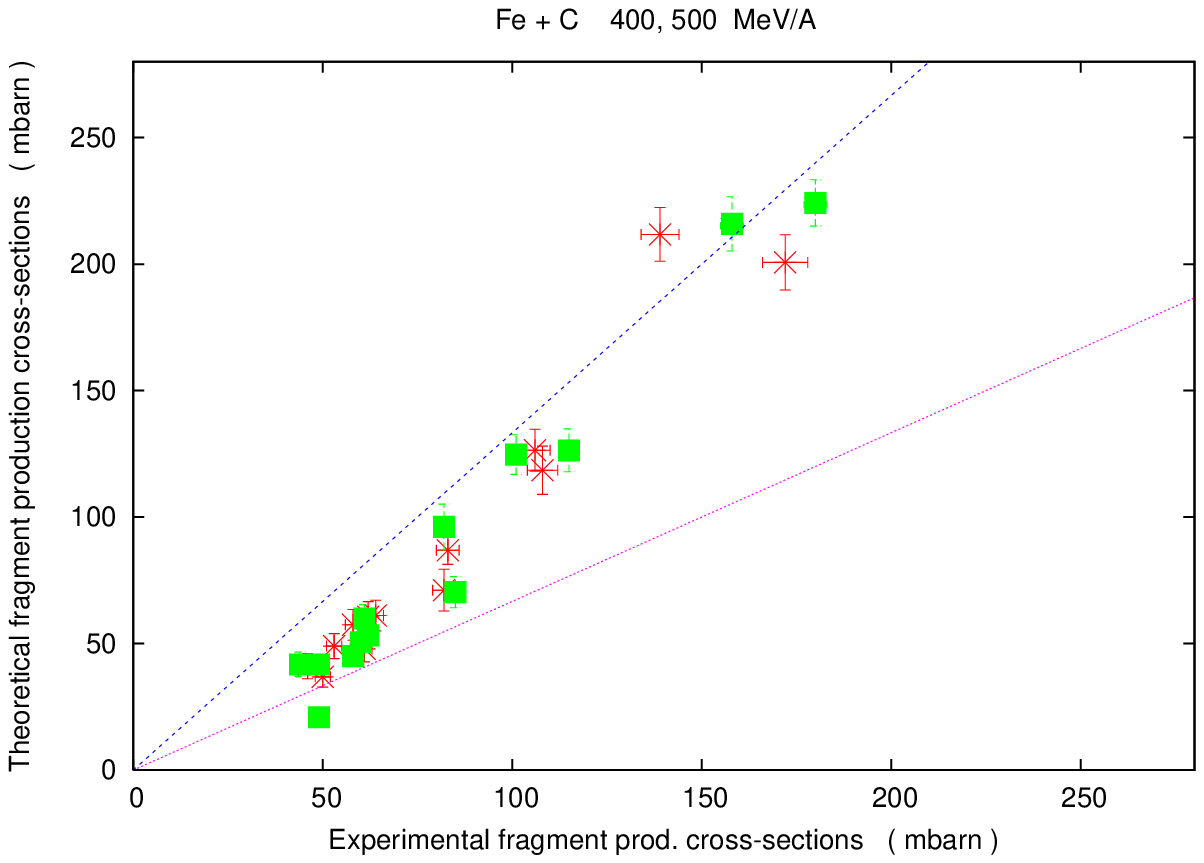}
\includegraphics[width=57mm]{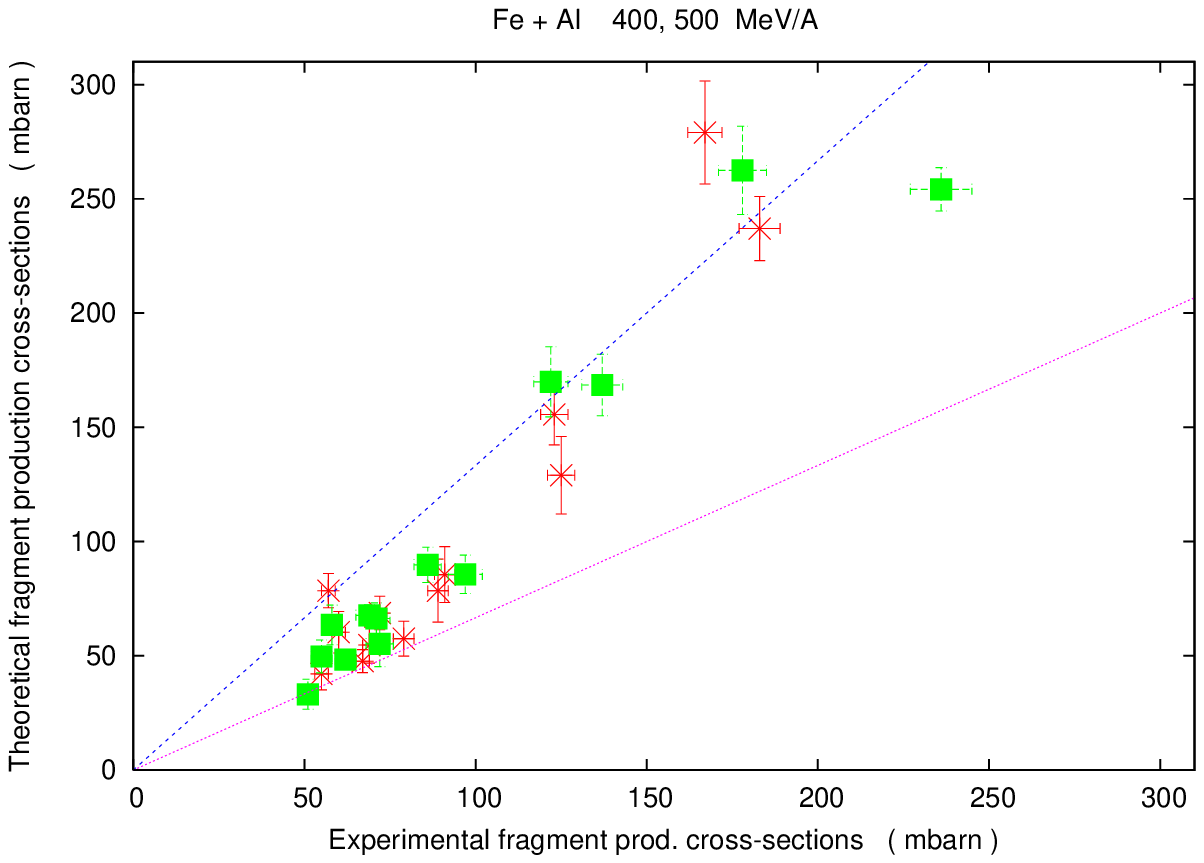}
\caption{\label{figfeh} Scatterplots of fragment production cross-sections 
predicted by QMD + FLUKA de-excitation (vertical axis) vs. experimental 
cross-sections~\cite{fragfrag} 
(horizontal axis) for Fe ions impinging on H (top left), 
C (top right) and Al (bottom) ions 
at  400 (crosses) and 500 (squares) MeV/A bombarding energies. 
The errorbars on the
theoretical results are sta\-ti\-sti\-cal only. The two lines respectively above
and below the diagonal correspond to deviations of the modelled results 
from the experimental ones by $\pm$ 33\%.}
\end{figure}

The same trend has also been observed in case of different projectiles.
The results for Fe + H, Fe + C and Fe + Al at 400 and 500 MeV/A
bombarding energies are plotted in the three panels of Fig.~\ref{figfeh}. 
The agreement between the results of our theoretical simulations
and the experimental data~\cite{fragfrag} is
well within 30 - 35 \% for most fragments. 
The QMD + FLUKA interface gives more reliable results in case
of heavy targets than in case of the hydrogen one. 
In case of the H target, the largest
discrepancies have been obtained 
for the lightest fragments (Z = 12 - 16) detected
by the experiment (the threshold for charge identification in these
experiments was Z~=~12) at the lowest bombarding energy, whereas a better 
agreement between theory and experiment is observed
for the Z = 22 - 25 fragment production cross-sections. 
In case of C and Al targets, the best agreement is observed for the 
Z = 14 - 21 fragments, whereas the largest discrepancies are
seen for the Z = 24 fragment cross-section, overpredicted by the theory. 
The question
about the dependence of these results on the target initial configuration
need further investigation. Up to now, we use only a few selected projectile
and target ion configurations 
to perform our simulations, giving rise to different
events by rotating in a random way the coordinates of these nuclei. The
simulations can be carried out as well by choosing a larger number of initial 
configurations (always satisfying some criteria concerning mean
square radius, binding energies, etc.) and one can study if this affects 
the results obtained for the most peripheral collisions. 
In fact, the most peripheral 
collisions are expected to be more sensitive than the central ones
to properties of the initial ion configurations such as 
root mean square radius oscillations.
  
Further effects are still lacking in our and in other QMD models.
Among the others, we mention shell effects, a spin-orbit term 
in the Hamiltonian, ground state deformation of initial nuclei.
In particular,
accounting for shape deformation can be important for very heavy systems, since
fusion/fission reactions are sensitive to the geometry of the collision and the relative orientation of the deformed nuclei. While the last effects 
can indeed be included by means of a proper choice of the initial nuclear 
states, introducing shell effects into a transport model is still a task 
of great difficulty~\cite{giant}. 

Furthermore, a proper treatment of antisymmetrization is still lacking in our model. However, it has already been included in some antisymmetrized advanced versions of QMD like AMD~\cite{amd} and FMD. These ones are slower, and have mainly been  used to study nuclear structure properties, especially for light nuclei, 
even if several examples of 
simulation of nuclear reactions with AMD have recently been produced.

\section{Isoscaling observables}
\label{iso}

\begin{figure}[h]\centering
\includegraphics[width=57mm]{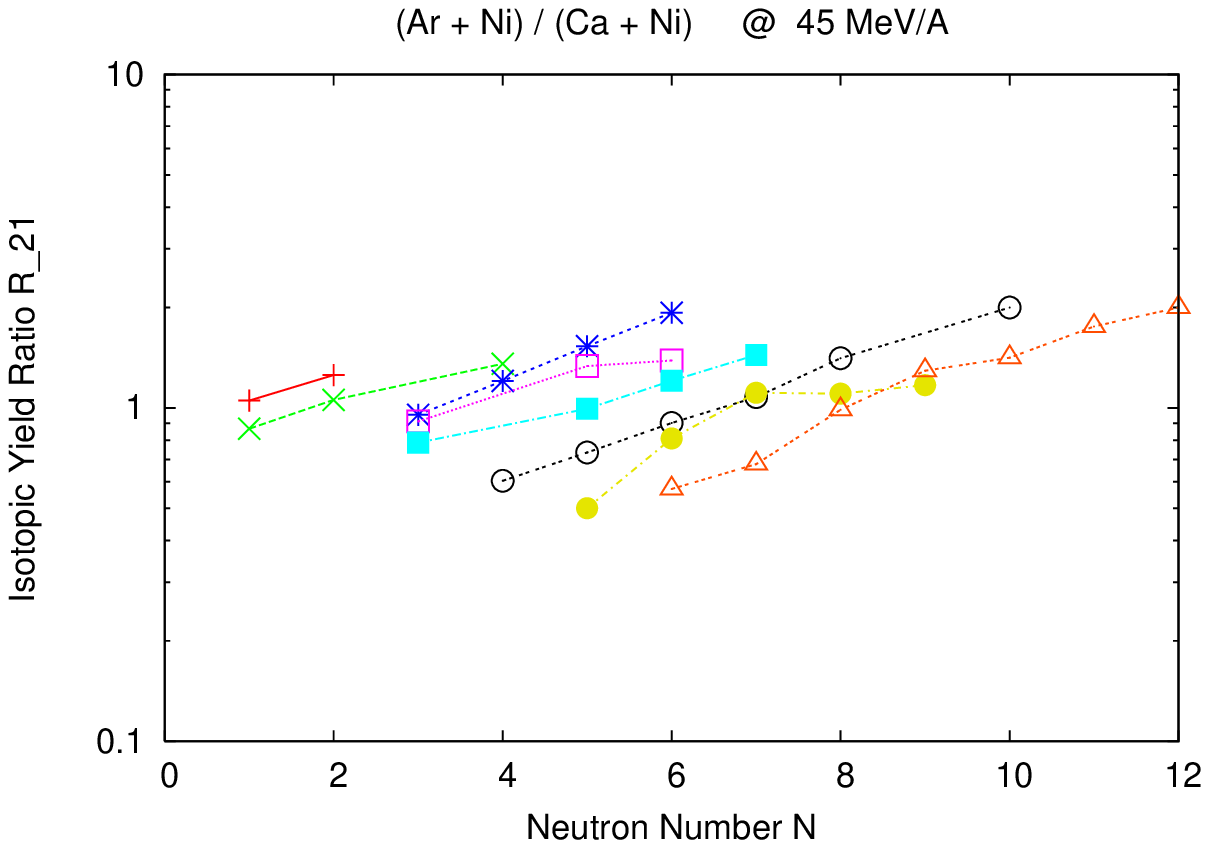}
\includegraphics[width=57mm]{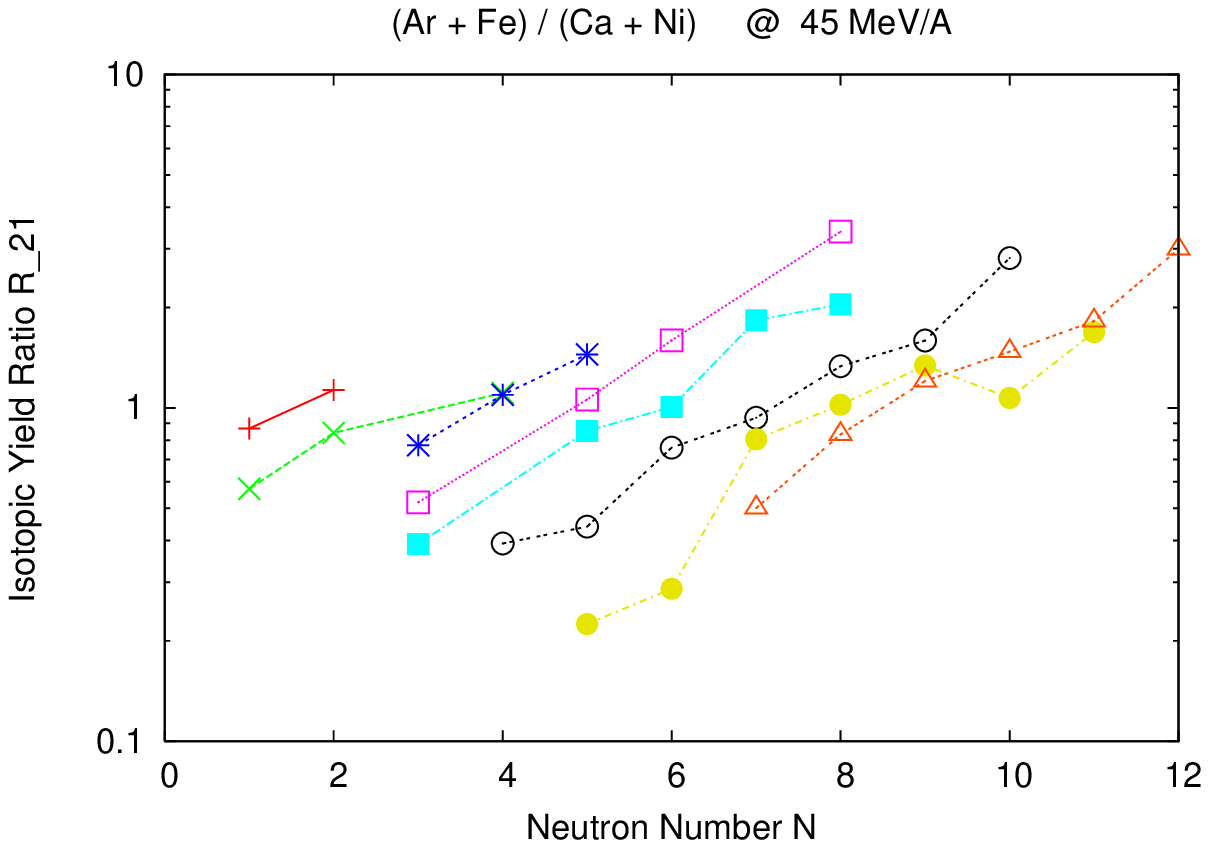}
\caption{\label{r21zeta} Fragment isotopic yield ratios obtained at 
the end of our simulations by
QMD + FLUKA de-excitation, as a function of N for a 45 MeV/A beam energy. The 
ratios for the Ar~+~Ni~/~Ca~+~Ni and Ar + Fe / Ca + Ni couple of reactions are
shown in the left and in the right panel, respectively. 
The different symbols correspond to
different values of Z in the Z~=~1~-~8 (from the left to the right
of each panel) range.}
\end{figure}

\begin{figure}[h]\centering
\includegraphics[width=116mm]{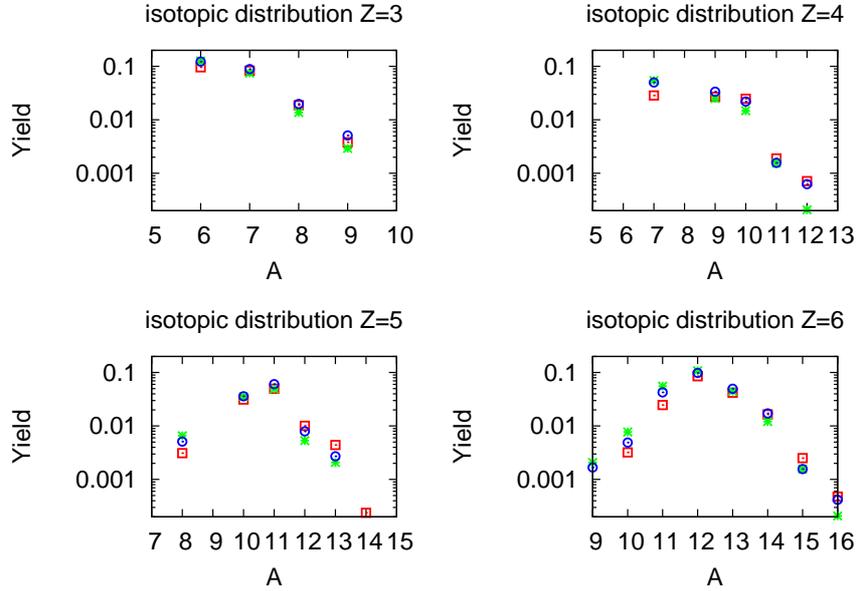}
\caption{\label{fragmentiso} Fragment isotopic distributions for
Li, Be, B and C produced by Ca + Ni (stars), Ar + Ni (circles) and
Ar + Fe (squares) reactions at a 45 MeV/A bombarding energy, at the
end of our simulations by QMD + FLUKA de-excitation.} 
\end{figure}

The isoscaling technique allows to obtain useful information concerning
the symmetry energy of isospin asymmetric nuclear matter, one of the quantities most uncertain
and most di\-scus\-sed. 
This information is important to understand the mechanisms
which regulate neutron star formation and structure, collapses of
massive stars and supernova explosions.
Both in theoretical simulations of nuclear collisions and in experiments 
the isotopic yield ratio from two reactions with 
similar total sizes and temperatures, 
but different isospin asymmetries 
$(\mathrm{N}-\mathrm{Z})/(\mathrm{N}+\mathrm{Z})$, 
has been found to
follow an exponential law~\cite{tsang,gulmi}:
\begin{eqnarray}
\frac{Y_{(2)}(\mathrm{N},\mathrm{Z})}{Y_{(1)}(\mathrm{N},\mathrm{Z})} = \exp(\alpha(\mathrm{Z}) \mathrm{N} + K(\mathrm{Z}))
\end{eqnarray}
where the index 2 refers to the reaction with the largest neutron
number (more asymmetric) and the index 1 refers to the reaction with
the largest proton number (more symmetric). 
This relation is valid at fixed $\mathrm{Z}$, limiting
to the isotopes $(\mathrm{N}, \mathrm{Z})$ 
for which the isotopic yield distributions 
can be approximated by gaussians. 
This means that the logarithm of the yield ratio is linear in N, at fixed Z,
with a slope given by $\alpha(\mathrm{Z})$. 
An analogous relation is valid at fixed N, 
meaning that the yield ratio is also linear in Z. 
This result is known as the isoscaling phenomenon and has been 
extensively studied~\cite{daniele, yenne}. Among the others, systematic 
experiments were performed at the Cyclotron Institute of TAMU 
and at the NSCL of MSU, analyzing 
the systems Ca~+~Ni, Ar~+~Ni, Ar~+~Fe at 25, 33, 45 and 53 MeV/A bombarding 
energies~\cite{yennello}. The experimental isotopic yield ratios 
for light fragments show a linear
dependence on N at fixed Z in the logarithmic plane, 
with $\alpha(\mathrm{Z})$ well 
approximated by a constant $\alpha$ 
for each fixed bombarding energy and couple $(1,2)$ of reactions.
The results have been interpreted by the authors of~\cite{yennello} 
both in the framework of the AMD dy\-na\-mi\-cal 
model and in the framework of a 
statistical model.
We have made simulation of the Ca + Ni, Ar + Fe and  Ar + Ni reactions 
at 45 MeV/A using the QMD model interfaced to the FLUKA de-excitation 
module. 
The isospin compositions N/Z for Ca~+~Ni, Ar~+~Ni and Ar + Fe amount
to $\approx$ 1.04, 1.13 and 1.18, respectively. 
Our results for the isotopic yield ratios at the end of
the whole simulation are plotted in Fig.~\ref{r21zeta}.
In each panel, also the results for the lightest fragments
(Z = 1, Z = 2) are included. 
It is apparent from our simulation that the linear dependence on N 
of the fragment yield ratios at fixed Z 
is a good approximation of the results in the logarithmic plane. 
Slight deviations from this linear dependence can be
due to low statistics ($\approx$ 
10000 events distributed over all possible impact
parameters were si\-mu\-la\-ted for each reaction).
By comparing the left and the right panel it is evident that larger slopes
are found for the yield ratios from the couple of reactions 
with larger dif\-fe\-ren\-ce
in the isospin compositions
($\mathrm{N}_2/\mathrm{Z}_2 - \mathrm{N}_1/\mathrm{Z}_1$). 
This behaviour is expected and in qualitative agreement with the experimental
observations. However, we observe values of $\alpha$ which differ with Z. 
We calculate an average $\alpha$ value using the $\alpha(\mathrm{Z})$ values in
the Z = 3 - 7 range. We obtain average $\alpha$ values $\approx$ 0.18 
and $\approx$ 0.31, 
for the Ar + Ni / Ca + Ni and Ar + Fe / Ca + Ni ratios, respectively.
These values are larger that the ones observed experimentally.  A reason
of these di\-scre\-pan\-cies is the fact that only the
fragments detected around a $44^{\mathrm{o}}$
angle were selected for performing 
the experimental isoscaling analysis, 
whereas we consider all fragments, independent of the
emission angles. This means that in the experiment the largest contribution
to the selected fragments comes probably from central collisions, and these
fragments are expected to include nucleons from both  the projectile ion 
and the target one,  whereas in our inclusive simulations 
IMF from projectile-like or target-like residues de-excitation are included.  
A preliminary analysis on our simulated events show that the $\alpha$
value is indeed affected by the choice of the impact pa\-ra\-me\-ter, and  
decreases significantly when selecting only the most central events. 
This behaviour is in qualitative agreement with Ref.~\cite{tsang}, which
claims that isoscaling is observed for a variety of
reaction mechanisms, from multifragmentation to
evaporation and DIS events, with different slopes in the logarithmic plane.  
Furthermore, we can observe that the results of our
analysis are quite sensitive to the number of isotopes included in the
linear fit at fixed Z or, in other words, 
to the goodness of the gaussian approximation to the
fragment isotopic distributions. The last ones are plotted for fragments
with Z = 3 - 6 in Fig.~\ref{fragmentiso}. 
In each panel the isotopic distributions
from the three considered reactions are plotted together.
The isotopic distributions for the 
emission of light fragments turn out to
show me\-mo\-ry of the initial isospin content of
the system of re\-ac\-ting ions. When comparing the results
for the three reactions, it is apparent that the reaction with the 
smallest asymmetry (or the smallest N/Z value) gives rise to isotopic 
distributions with larger tails at smaller N, whereas the one with 
the largest asymmetry (or the largest N/Z value) gives rise to larger 
tails at larger N. This result is in qualitative 
agreement with experimental observations. 
This behaviour is expected to become even more evident at lower 
bombarding energies.

When considering the free particles emitted in the
fast stage of the collision, our QMD simulation
suggests that the average yield of emitted protons is larger 
than the one of emitted neutrons in the Ca + Ni reaction, which has 
the lo\-west N/Z value. The other reactions exhibit the opposite trend 
for the collisions at low impact parameters. 
This behaviour is due to Coulomb force. In particular, 
according to our simulations, 
at pre-equilibrium, the yield of protons $\mathrm{Y_p}$ emitted in Ca + Ni 
central collisions is a fraction 
$\approx$ 20\% larger than the yield of emitted neutrons $\mathrm{Y_n}$, 
whereas
$\mathrm{Y_p}$ is $\approx$ 10\% and 15\% smaller than $\mathrm{Y_n}$ 
for Ar + Ni and Ar + Fe 
central collisions, respectively. 
   
We have also investigated up to which extent the values of $\alpha$ are
sensitive to the use of the final fragments (from the full simulation) or
the use of the primary excited fragments in performing the yield ratio
analysis. 
We have found that, for the reactions considered in this study,
the average value of $\alpha$ at the end of the overlapping stage 
of the collisions described by QMD, denoted 
by $\alpha_{hot}$, can be larger than the value of $\alpha$ 
at the end of the full simulation including de-excitation
effects by no more than $\approx$ 20\%.

Finally, we calculate the fragment asymmetry $\mathrm{(Z/A)_{liq}}^2$ 
of the liquid phase for the three reactions at t = 250 fm/c. 
According to Ref.~\cite{tsang}, the liquid phase is 
defined as composed by all fragments with A $>$ 4. 
For central collisions in each of the three systems
the ave\-ra\-ge $\mathrm{(Z/A)_{liq}}^2$ at the end of the pre-equilibrium
stage turns out to be 
lower then the corresponding value at t = 0, as already found for
the same reactions at a lower energy by~\cite{yennello},
with a difference between the values at different time 
more evident for the most symmetric reaction (Ca + Ni).
The dependence of the average $\mathrm{(Z/A)_{liq}}^2$ 
on time and impact parameter is under further investigation.

\section{Conclusions and practical applications}
\label{conclu}

A few examples of the application of the QMD code developed in Milano
and interfaced to the FLUKA de-excitation module have been produced, 
in the simulation of heavy-ion collisions 
both at a low bombarding energy (E $\approx$ 50 MeV/A) 
and at a few hundreds MeV/A beam energies, showing results which are
in agreement, at least from a qualitative point of view, with experimental
data and/or previous studies. 
The studies at low bombarding energies are mainly
driven by theoretical issues and are aimed at gaining a better understanding
of the role of isospin in multifragmentation processes and in astrophysical
mechanisms crucial to determine the evolution of supernovae and neutron stars,
and can give insights on the optimization of the production of rare 
isotopes far from stability.
On the other hand, the possibility to extend simulations by the same
model even at higher bombarding energies allows to cover application purposes 
too. 

Among the practical application of QMD models we mention hadrontherapy,
civil aviation and space radiation protection, and single-event effects in
microelectronics.
As far as hadrontherapy is concerned, ion beams (in particular C beams) are 
used to produce Complex Lesions, i.e. multiple breaks of the DNA double 
strands 
contained in cells. This effect can be exploited in order to kill tumours. 
This application requires the integration 
of nuclear models with radiobiological models.
In particular, the ion biological effectivenesses turns out to be 
larger than the  
X-ray and proton ones.  
A few facilities built on the basis of 
these principles already exist (e.g. in Japan and in Germany, at GSI and in Heidelberg) or are under construction (e.g. the CNAO in Pavia, Italy).
At higher energies, crews in space missions and in air-flight are exposed 
to Galactic Cosmic Rays (GCR). Spacecraft and aeroplanes have to be carefully 
designed in order to minimize GCR interaction effects~\cite{cesca}. 
Furthermore, functional upset of microelectronic memory devices 
(silicon chips) in space missions can also occur and 
deserves particular attention~\cite{watanabe}. 
This phenomen can even
take place at sea level, due to atmospheric neutrons (produced in the interactions of Cosmic Rays with air) in the 50 - 1000 MeV energy range.

\section*{Acknowledgments} 

The author is grateful to D. Satoh for useful discussions and for 
providing ex\-pe\-ri\-men\-tal 
data concerning neutron emission from heavy-ion
reactions in thick targets.
Collaboration with F. Ballarini, G. Battistoni, F. Cerutti, A. Fass\`o
A. Ferrari,  E. Gadioli, A. Ottolenghi, L.S. Pinsky, J. Ranft, P.R. Sala
is acknowledged. The FLUKA code is copyrighted by INFN-CERN and is
available on the web at \texttt{http://www.fluka.org}. 
This work is supported by the University of Milano.

\end{document}